\documentclass{ifacconf}
\makeatletter
\let\algorithm\@undefined
\let\endalgorithm\@undefined
\makeatother
\usepackage[ruled,vlined,linesnumbered]{algorithm2e}
\usepackage{graphicx,siunitx,balance}      
\usepackage{natbib}        
\usepackage{amssymb,mathtools,theorem,psfrag,comment}
\usepackage{amsmath,color}

\newtheorem{theorem}{Theorem}

\newtheorem{proposition}{Proposition}

\newtheorem{assumption}{Assumption}
\newtheorem{definition}{Definition}
\newtheorem{lemma}{Lemma}
\newtheorem{problem}{Problem}
\newtheorem{remark}{Remark}
\newtheorem{property}{Property}

\newcommand*{\QEDA}{\hfill\ensuremath{\blacksquare}}%
\DeclareMathOperator{\q}{q}
\DeclareMathOperator{\He}{He}
\DeclareMathOperator{\dom}{dom}
\DeclareMathOperator{\Sign}{sign}

\DeclareMathOperator{\rge}{range}

\DeclareMathOperator{\specr}{spr}

\DeclareMathOperator{\maximize}{maximize}
\DeclareMathOperator{\minimize}{minimize}
\newcommand{\R}{\mathbb{R}}
\begin{document}
\begin{frontmatter}

\title{Sampled-data control design for systems with quantized actuators\\ (Extended Version)}
\thanks{This work has been supported in part by ANR via project HANDY, number ANR-18-CE40-0010.}
\thanks{\textcolor{blue}{This file contains fixes to some (minor) typos appearing in the published conference proceedings. Typos are in blue font.}}

\author[UNIPG]{Francesco Ferrante} 
\author[LAAS]{Sophie Tarbouriech}

\address[UNIPG]{Department of Engineering, University of Perugia, Perugia, Italy. francesco.ferrante@unipg.it}
\address[LAAS]{LAAS-CNRS, Universit\'e de Toulouse, CNRS, Toulouse, France. tarbour@laas.fr}

\begin{abstract}                
This paper deals with the problem of designing a sampled-data state feedback control law  for continuous-time linear control systems subject to uniform input quantization. The sampled-data state feedback is designed in order to ensure uniform global asymptotic stability (UGAS) of an attractor surrounding the origin. The closed-loop system is rewritten as a hybrid dynamical system. To do this, an auxiliary clock variable triggering the occurrence of sampling events is introduced. A numerically tractable algorithm with feasibility guarantees, based on concave-convex decomposition, is then proposed allowing to minimize the size of the attractor. Theoretical results are illustrated in a numerical example.\end{abstract} 

\end{frontmatter}

\section{Introduction}
The design of a new generation of control systems combining physical interactions with computational and communication abilities presents strong advantages in terms of scalability,  maintenance, and computational resources, for example for transportation systems, autonomous robotics, and energy delivery systems (\cite{mur:ast:boy:bro:ste/repor2002}). This paper focuses on one of these digital elements, namely, quantization, which is typically used to reduce data traffic load in communication channels to comply with limited bandwidth constraints. As a side effect, quantization induces a nonlinear behavior into the closed loop. See, for example, \cite{delchamps1990stabilizing}, \cite{ceragioli2011discontinuities},  \cite{lib/book2003}, \cite{tarbouriech2011control}. Most of the works dealing with quantized control systems focus either on continuous-time systems (\cite{RogerBrockett}, \cite{DanielLiberzon}, \cite{EmiliaFridman2},  \cite{ferrante2015stabilization}, \cite{ferrante2020sensor}), or on discrete-time systems,
(\cite{BrunoPicasso}, \cite{MinyueFu-ieee09}, \cite{campos2018stabilisation}, \cite{ich:saw:tar/ijrnc2018}). However, in many applications the actual setting consists of a physical plant that is controlled via a digital controller. Because the dynamics of the physical plants are naturally described by continuous-time systems, the resulting closed-loop system turns out to be a sampled-data system. Sampled-data systems have been extensively studied in the last twenty years; see \cite{hetel2017recent} for a recent survey on sampled-data systems. One might argue that the impact of quantization in sampled-data systems may be studied by relying on a discretized model of the plant. However, by pursuing this approach the analysis of the intersample behavior is completely lost. 

Despite the fact that quantization and sampling naturally coexist, surprisingly not much attention has been devoted to the control design of quantized sampled-data systems. The interplay of sampling and quantization has been analyzed in \cite{nesic2009unified}. Therein, a unified framework for controller design in the presence of  quantization and time scheduling based on an emulation approach is investigated. Stabilization of sampled-data systems with stochastic sampling times in the presence of logarithmic input quantization has been addressed in \cite{shen2017quantized} by relying on a discretized model of the closed-loop system. 
Recently, robust sampled-data stabilization in the presence of input/state quantization and uncertainties for a rather general class nonlinear systems has been studied in \cite{di2021robust}, where the authors provide sufficient conditions for semiglobal practical asymptotic stability in the presence of fast sampling and accurate quantization.


In this paper, we consider a setup in which a linear plant is controlled via a periodic uniformly quantized sampled-data state feedback controller. In contrast with \cite{shen2017quantized} and with the objective of fully characterizing the intersample behavior of the closed-loop system, we model the closed loop as a hybrid dynamical system in the framework of \cite{goebel2012hybrid}. Within this setting, we provide sufficient conditions for uniform global asymptotic stability of a compact set containing the origin of the plant substate. These conditions are later used to devise a design algorithm for the controller to mitigate the impact of input quantization on the closed-loop state. It is noteworthy to mention that, as opposed to  \cite{di2021robust}, we assume that the quantization error bound and the sampling time are prescribed and cannot be selected by the designer. In this sense, the approach we propose is fully constructive.  

The remainder of the paper is organized as follows. Section~\ref{sec:pb} formalizes the stabilization problem we solve. Section~\ref{sec:SuffCond} provides sufficient conditions in the form of matrix inequalities for closed-loop stability. Section~\ref{sec:CompDesign} presents an optimal design algorithm for the controller. The applicability of the proposed results is illustrated in Section~\ref{sec:Ex} via a numerical example.
    
\subsection*{Notation}
The symbol $\mathbb{N}$ denotes the set of nonnegative integers, $\mathbb{N}_{>0}$ is the set of positive integers, $\R$ represents the set of real numbers, $\R^n$ is the $n$-dimensional Euclidean space, $\R^n_{>0}$ is the positive (open) orthant in $\R^n$, $\R^{n\times m}$ represents the set of $n\times m$ real matrices, $\mathbb{Z}$ is the set of relative integers,
and $\mathbb{B}$ is the closed unitary ball in the Euclidean norm. Given $\delta>0$, $\delta\mathbb{Z}\coloneqq\{\delta k\colon k\in\mathbb{Z}\}$.
The symbol $\mathbb{S}_{+}^n$ ($\mathbb{S}_{++}^n$) stands for the set of $n\times n$ symmetric positive semidefinite (definite) matrices, 
 $\mathbb{D}_+^{n}$  ($\mathbb{D}_{++}^{n}$) denotes the set of $n\times n$ diagonal positive semidefinite (definite)  matrices. For a vector $x\in\R^n$ (a matrix $A\in\R^{n\times m}$) $x^\top$ ($A^\top$) denotes the transpose of $x$ ($A$). The notation $\Vert A\Vert$ indicates the spectral norm of the matrix $A$.  The spectral radius of the matrix $A$ is denoted by $\specr(A)$. Given $A\in\R^{n\times n}$, $\He (A)=A+A^\top$. The symbol $A\preceq 0$ ($A\prec 0$) stands for  negative (semi) definiteness of the symmetric matrix $A$. The symbol $\star$ stands for symmetric blocks in symmetric matrices. Given a symmetric matrix $A$, 
$\lambda_{\max}(A)$ and $\lambda_{\min}(A)$ stand, respectively, for the largest and smallest eigenvalue of $A$. Given $A\in\R^{n\times n}$, the notation $\mathcal{E}(A)=\{x\in\R^n\colon x^\top A x\leq 1 \}$ is used. The function $\Sign\colon\R\to\{-1, 1\}$ is defined for all $x\in\R$ as follows: $\Sign(x)=1$ if $x\geq 0$ and $-1$ otherwise. The symbol $\vert x\vert_{S}\coloneqq\displaystyle\inf_{y\in S}\vert x-y\vert$ denotes the distance of the point $x\in\R^n$ to the nonempty set $S\subset \R^n$. The symbol $\lfloor x\rfloor$ indicates the floor of the real number $x$. The notation $L_V(c)$ stands for the $c$-sublevel set of the function $V$, i.e., $L_V(c)\coloneqq\{x\in\dom V\colon V(x)\leq c\}$. Let $f\colon \mathcal{X}\to\mathcal{Y}$, and $x\in\mathcal{X}$, we denote by $Df(x)\colon\mathcal{S}\to\mathcal{Y}$ the differential of $f$ at $x$. 
The symbol $\rge f$ indicates the image of the function $f$. The notation $\overline{S}$ indicates the closure of the set $S$. 
\section{Problem Statement and Modeling}
\label{sec:pb}
\subsection{Problem setup}
We consider the following continuous-time plant with quantized actuation:
\begin{equation}\label{1}
\begin{aligned}
 & \dot{x}_p=A_px_p+B_p\q_\Delta(u_p)
 \end{aligned}
\end{equation}
where $x_p\in\mathbb{R}^{n_p}$ and $u_p\in\mathbb{R}^{n_u}$ are, respectively, the plant state and control input. Matrices $A_p\in\R^{n_p\times n_p}$ and $B_p\in\R^{n_p\times n_u}$ are assumed to be known. The function $\q_\Delta$ is the so-called \emph{uniform quantizer}, which is defined next:
$$
u\mapsto q_\Delta(u)=(q_{\delta_1}(u_{1}), q_{\delta_2}(u_{2}), \dots, q_{\delta_{n_u}}(u_{n_u}))\in \Delta\mathbb{Z}^{n_u}
$$
where $\Delta\coloneqq(\delta_1, \delta_2,\dots, \delta_{n_u})\in\R_{>0}^{n_u}$ represents 
the vector of quantization levels of each channel, $\Delta\mathbb{Z}^{n_u}\coloneqq\bigtimes_{i=1}^{n_u}\delta_i\mathbb{Z}$, and, for all $v_i\in\R$, $\delta_i\in\R_{>0}$ 
$$
q_{\delta_i}(v_i)\coloneqq \delta  \Sign(v_i)\Bigl\lfloor{\frac{\vert v_i\vert}{\delta_i}\Bigr\rfloor}.
$$
We consider a scenario in which the state of plant \eqref{1} is controlled via the following sampled-data state feedback controller:
\begin{equation}\label{2}
\begin{aligned}
 &u_c(t_k)=K_1 x_p(t_{k})+K_2 u_c(t_{k-1})
 \end{aligned}  
\end{equation}
where $\textcolor{blue}{t_0=0}$ and $t_1, t_2, \dots$ denotes the sampling instances and $K_1\in\R^{n_u\times n_p}$ and $K_2\in\R^{n_u\times n_u}$ are two gains defining the controller dynamics to be designed. Within this setting, our goal is to design controller \eqref{2} ensuring closed-loop stability. 

Throughout the paper, we assume that the sequence of sampling times is periodic, formally the following assumption is enforced:
\begin{assumption}
\label{assu:sampling}
For the sequence $\{t_k\}_{\textcolor{blue}{1}}^\infty$, there exist $T>0$ such that
$$
\begin{aligned}
&\textcolor{blue}{0}\leq t_{1}\leq T\\
&t_{k+1}-t_{k}=T, &\forall k\in\mathbb{N}_{>0}.
\end{aligned}
$$
\hfill$\triangle$
\end{assumption}
Assumption~\ref{assu:sampling} ensures that the sequence of sampling times is strictly increasing and unbounded. In particular, assuming $T>0$ rules out the existence of accumulation points in $\{t_k\}_{\textcolor{blue}{1}}^\infty$. 

The interconnection of the continuous-time plant \eqref{1} and the sampled-data controller \eqref{2} is realized via the following ``time-triggered'' zero-order-holder device:
\begin{equation}
\left\{\begin{aligned}
&\dot{\chi}(t)=0&\,\,\text{when}\,\, t\notin\{t_k\}_{\textcolor{blue}{1}}^\infty\\
&\chi(t)=v(t)&\,\,\text{when}\,\, t\in\{t_k\}_{\textcolor{blue}{1}}^\infty,
\end{aligned}\right.
\label{eq:ZOH}
\end{equation}
by setting $\chi= u_p$ and $v=u_c$.
\subsection{Hybrid modeling}
Due to the interplay of continuous-time dynamics and discrete-time updates in the closed-loop system  \eqref{1}-\eqref{2}-\eqref{eq:ZOH}, we recast it as a hybrid dynamical system in the framework in \cite{goebel2012hybrid}. To this end, we introduce an auxiliary clock variable $\tau$ triggering the occurrence of sampling events. In particular, we let $\tau$ increase as long as its value belongs to the interval $[0, T]$ and trigger a jump whenever $\tau=T$. After the jump, the value of $\tau$ is reset to $0$. This leads to the following dynamics for $\tau$:
 \begin{equation}
 \label{eq:clock}
 \left\{\begin{aligned}
 &\dot{\tau}=1,&\,\,\tau\in [0, T]\\
 &\tau^+=0,&\,\,\tau=T. 
 \end{aligned}\right.
 \end{equation}
Having introduced this clock enables us to model the closed-loop system  \eqref{1}-\eqref{2}-\eqref{eq:ZOH} as the following hybrid system with state $x\coloneqq (x_p, \chi, \tau)=\colon (\xi, \tau)\in\R^{n_p+n_u+1}$:
\begin{subequations}
\label{eq:HybCL}
\begin{equation}
\mathcal{H}_{cl}\left\{\begin{aligned}
&\begin{bmatrix}
\dot{\xi}\\
\dot{\tau}
\end{bmatrix}=\underbrace{\begin{bmatrix}
A_{cl}\xi\\
1
\end{bmatrix}}_{f(x)},&x\in\mathcal{C}
\\[2mm]
&\begin{bmatrix}
\xi^+\\
\tau^+
\end{bmatrix}=\underbrace{
\begin{bmatrix}
(G_{cl}+J_{cl}K)\xi+J_{cl}\psi_\Delta(K\xi)\\
0
\end{bmatrix}}_{g(x)},&x\in\mathcal{D}
\end{aligned}\right.
\end{equation}
where: 
\begin{equation}
\begin{aligned}
\label{eq:flow_jump_matrices}
&A_{cl}\coloneqq\begin{bmatrix}
A_p&B_p\\
0&0
\end{bmatrix}, G_{cl}\coloneqq\begin{bmatrix}
I&0\\
0&0
\end{bmatrix}, J_{cl}\coloneqq\begin{bmatrix}0\\I\end{bmatrix}, \\
&K\coloneqq\begin{bmatrix}K_1&K_2\end{bmatrix},
\end{aligned}
\end{equation}

\begin{equation}
\label{eq:flow_jump_sets}
\begin{aligned}
&\mathcal{C}\coloneqq\R^{n_p}\times\Delta\mathbb{Z}^{n_u}\times [0, T],\\
&\mathcal{D}\coloneqq\R^{n_p}\times\Delta\mathbb{Z}^{n_u}\times \{T\},
\end{aligned}
\end{equation}
and
\begin{equation}
\label{eq:Psi}
\begin{aligned}
&\psi_\Delta(u)\coloneqq \q_\Delta(u)-u, \quad\forall u\in\R^{n_u}.
\end{aligned}
\end{equation}
\end{subequations}
\medskip

\begin{remark}
As opposed to the original model, in the proposed hybrid model \eqref{eq:HybCL} the quantizer is placed in the jump map, thereby rendering the flow map affine. Notice that, due to the specific selection of the flow and jumps sets, this operation does not induce any modification of the closed-loop system behavior. 
\hfill$\diamond$
\end{remark}
\subsection{Basic properties of $\mathcal{H}_{cl}$ and Krasovskii regularization}
In this subsection, we illustrate some structural properties of hybrid model $\mathcal{H}_{cl}$. 
\begin{property}
\label{prop:exi}
For hybrid system $\mathcal{H}_{cl}$, the following properties holds:
\begin{itemize}
\item[$(i)$] For any $\xi\in\mathcal{C}\cup\mathcal{D}$, there exists a nontrivial solution $\phi$ to 
$\mathcal{H}_{cl}$ with $\phi(0, 0)=\xi$;
\item[$(ii)$] Any maximal solution to $\mathcal{H}_{cl}$ is complete;
\item[$(iii)$] Any complete solution $\phi$ is $t$-complete, i.e., $\sup_t\dom\phi=\infty$. 
\end{itemize}
\end{property} 
\begin{pf}
The proof hinges upon direct application of \cite[Proposition 2.10, page 33]{goebel2012hybrid}. In particular, the viability condition (VC) in \cite[Proposition 2.10, page 33]{goebel2012hybrid} trivially holds on $\mathcal{C}\setminus\mathcal{D}=\R^{n_p}\times\R^{n_u}\times [0, T)$
due to the flow dynamics being affine and the dynamics of $\tau$ allowing flows when $\tau\in [0, T)$.  
This yields item $(i)$. To show item $(ii)$, still from \cite[Proposition 2.10, page 33]{goebel2012hybrid},
since no finite escape times are possible and $g(\mathcal{D})\subset\mathcal{C}$, item $(ii)$ follows directly. To conclude the proof, observe that $g(\mathcal{D})\subset(\mathcal{C}\setminus\mathcal{D})$. Moreover, the timer dynamics \eqref{eq:clock} ensure that the length of any flow interval is lower bounded by $T$. Thus, item $(iii)$ follows.
\QEDA
\end{pf}

The above result shows that hybrid system $\mathcal{H}_{cl}$ is well behaved in terms of existence and completeness of (maximal) solutions.  However,  the discontinuity of the map $\psi_\Delta$ prevents $
\mathcal{H}_{cl}$ from fulfilling the so-called \emph{hybrid basic conditions}; see \cite[Assumption 6.5, page 
120]{goebel2012hybrid}. Hybrid systems not fulfilling the hybrid basic conditions may be overly sensitive to 
small vanishing noise. Moreover, the fulfillment of the hybrid basic conditions is a key ingredient for the 
application of several important results. To overcome this problem, in this paper we consider the so-called \emph{Krasovskii regularization} of $\mathcal{H}_{cl}$; see \cite[Definition 4.13, page 81]{goebel2012hybrid}.  
This regularization basically corresponds to the ``smallest'' hybrid dynamical systems containing  $\mathcal{H}_{cl}$ fulfilling the hybrid basic conditions. In the specific case of hybrid system $\mathcal{H}_{cl}$, such a regularization writes as:
\begin{equation}
\label{eq:H_cl}
\widehat{\mathcal{H}}_{cl}\left\{\begin{aligned}
&\dot{x}=f(x), &x\in\mathcal{C}\\
&x^+\in G(x), &x\in\mathcal{D}
\end{aligned}\right.
\end{equation}
where for all $x\in\mathcal{D}$
$$
G(x)\coloneqq \bigcap_{\delta>0} \overline{g((x+\delta\mathbb{B})\cap\mathcal{D})}.
$$
Because $g$ is locally bounded on $\mathcal{D}$, similarly as in \cite[Lemma 1]{bacciotti2004stabilization}, it is easy to see that for all $x\in \mathcal{D}$:
$$
\begin{aligned}
G(x)=&\left\{v\in\R^{n_p+n_u+1}\colon\
 \exists \{x_k\}\subset\mathcal{D}\right.\\
&\qquad\qquad\left.\mbox{s.t.}\, x_k\to x, \textcolor{blue}{g(x_k)}\to v\right\}.
\end{aligned}
$$
Using the above equivalent rewriting of $G$, simple continuity arguments yield:
$$
G(x)=\begin{bmatrix}
(G_{cl}+J_{cl}K)\xi+J_{cl}\Phi(\xi)\\
0
\end{bmatrix}, \quad\forall x\in\mathcal{D}
$$
where for all $\xi\in\R^{n_p+n_u}$:
\begin{equation}
\label{eq:Phi}
\Phi(\xi)\coloneqq\bigcap_{\delta>0} \overline{\psi_\Delta(K(\xi+\delta\mathbb{B}))}.
\end{equation}
It is worth to observe that since $g(\mathcal{D})\subset G(\mathcal{D})$, solutions to hybrid system  \eqref{eq:HybCL} are solutions to \eqref{eq:H_cl}. Therefore, stability properties established for \eqref{eq:H_cl}, among others, carry over directly to \eqref{eq:HybCL}. Moreover, completeness of maximal solutions to \eqref{eq:H_cl} is still ensured.

To simplify the analysis of the closed-loop system, we over-approximate the set-valued map $\Phi$ via the following inclusion:
\begin{equation}
\label{eq:PSI_inc}
\Phi(\xi)\subset \Psi_\Delta(K\xi), \quad \forall \xi\in\R^{n_p+n_u}
\end{equation}
where $\Psi_\Delta$ is the Krasovksii regularization of the mapping $\psi_\Delta$, namely:
\begin{equation}
\label{eq:PSI}
\Psi_\Delta(u)\coloneqq\bigcap_{\delta>0} \overline{\psi_\Delta(u+\delta\mathbb{B})},\quad \forall u\in\R^{n_u}.
\end{equation}
In particular, \eqref{eq:PSI_inc} can be proven directly by relying on the continuity of the function $ \xi\mapsto K\xi$.
It is worth to observe that, since $g(\mathcal{D})\subset G(\mathcal{D})$, solutions to hybrid system  \eqref{eq:HybCL} are solutions to \eqref{eq:H_cl}. Therefore, stability properties established for \eqref{eq:H_cl}, among others, carry over directly to \eqref{eq:HybCL}. Moreover, due to 
$G(\mathcal{D})\subset(\mathcal{C}\setminus\mathcal{D})$, completeness and non-Zenoness of maximal solutions to \eqref{eq:H_cl} follows from similar arguments as in the proof of Property~\ref{prop:exi}.

At this stage, it is important to observe that, due to the presence of input quantization, generally asymptotic stabilization of the origin of the $\xi$-substate cannot be achieved. Indeed, roughly speaking, due to $\psi_\Delta$  being zero around zero, the closed-loop system behaves in open loop near the origin. Therefore, it is natural to turn the attention towards asymptotic stability of a ``possibly minimal'' compact set 
$\mathcal{A}\subset\R^{n_p+n_u+1}$ containing the origin of the $\xi$-substate.  This leads to the following statement for the stabilization problem we solve in this paper. 
\begin{problem}[Stabilization]
\label{pb:Design}
Design a gain $K$ such that there exists a compact set $\mathcal{A}\subset\R^{n_p+n_u+1}$ such that the following hold:
\begin{itemize}
\item[(a)]  $(\{0_{n_p}\}\times\{0_{n_u}\}\times [0, T])\subset{\cal A}$;\\
\item[(b)] ${\cal A}$ is uniformly globally asymptotically stable \emph{UGAS} for hybrid system \eqref{eq:H_cl}. 
\end{itemize}
\hfill $\diamond$
\end{problem}
\section{Sufficient conditions for stability}
\label{sec:SuffCond}
To address Problem~\ref{pb:Design}, we make use of the following sector conditions for the set-valued map $\Psi_\Delta$ in \eqref{eq:PSI}. 
\begin{lemma}
\label{lemm:Sector}
Let $S_1, S_2\in\mathbb{R}^{n_u\times n_u}$ be diagonal positive definite matrices. Then, for all $u\in\mathbb{R}^{n_u}$, $v\in\Psi_\Delta(u)$ the following inequalities hold:
\begin{align}
\label{eq:sec_1}
&v^\top S_1v-\Delta^\top S_1\Delta\leq0\\
\label{eq:sec_2}
&v^\top S_2(v+u)\leq 0.
\end{align}
\hfill$\diamond$
\end{lemma}
The proof Lemma~\ref{lemm:Sector} follows closely the proof of \cite[Lemma 4.2]{ferrante2020sensor}  and therefore it is omitted. 
\begin{theorem}
Let $T>0$ be given. Suppose that there exist $S_1, S_2\in\mathbb{D}_{++}^{n_u}$, $\varrho\in(0, 1)$, $P\in\mathbb{S}^{n_p+n_s}_{++}$, $K\in\mathbb{R}^{n_u\times (n_p+n_u)}$ such that
\begin{align}
\label{eq:trace_cond}
&\Delta^\top S_1\Delta-\varrho\leq 0,\\
&M\coloneqq\begin{bmatrix}
M_{11}&M_{12}\\
\star&M_{22}
\end{bmatrix}\prec 0
\label{eq:M_matrix}
\end{align}
where:
$$
\begin{aligned}
&M_{11}\coloneqq(G_{cl}+J_{cl}K)^\top \exp(A^\top_{cl}T) P\exp(A_{cl}T)(G_{cl}+J_{cl}K)\\
&\quad\quad+(\varrho-1)P,\\
&M_{12}\coloneqq (G_{cl}+J_{cl}K)^\top \exp(A^\top_{cl}T) P\exp(A_{cl}T)J_{cl}-K^\top S_2,\\
&M_{22}\coloneqq J_{cl}^\top \exp(A^\top_{cl}T) P\exp(A_{cl}T)J_{cl}-S_1-2S_2.
\end{aligned}
$$
Then, there exists $\sigma^\star>0$ small enough such that for all $\sigma\in (0, \sigma^\star]$ the set
\begin{equation}
\label{eq:Attract}
\mathcal{A}\coloneqq\{x\in\mathcal{C}\colon \xi^\top\exp(-\sigma\tau)\exp(A^\top_{cl}\tau) P\exp(A_{cl}\tau)\xi\leq 1\}
\end{equation}
is UGAS for the hybrid system \eqref{eq:H_cl}.
\label{theo:main}
\end{theorem}
\begin{pf}
Let for all $x=(\xi, \tau)\in\R^{n_p+n_u}\times\R$
$$
V(x)\coloneqq \exp(-\sigma\tau)\xi^\top \Sigma(\tau)^\top P\Sigma(\tau)\xi
$$
where $\sigma>0$ is selected later and, for the sake of writing, we set $\Sigma(\tau)\coloneqq\exp(A_{cl}(T-\tau))$.  The proof relies on Proposition~\ref{prop:gen_stab}. In particular, we show that under \eqref{eq:trace_cond} and  \eqref{eq:M_matrix}
all the assumptions in Proposition~\ref{prop:gen_stab} in Appendix~\ref{app:results} are satisfied with the above selection of $V$.

Observe that for all  $x=(\xi, \tau)\in\mathcal{C}=\mathcal{C}\cup\mathcal{D}\cup G(\mathcal{D})$ 
\begin{equation}
\label{eq:SandOmega}
\omega_1 \vert \xi\vert^2\leq V(x)\leq\omega_2 \vert \xi\vert^2
\end{equation}
where:
$$
\begin{aligned}
&\omega_1\coloneqq\min_{\tau\in[0, T]}\exp(-\sigma\tau)\lambda_{\min}\left(\Sigma(\tau)^\top P\Sigma(\tau)\right)\\
&\omega_2\coloneqq\max_{\tau\in[0, T]}\exp(-\sigma\tau)\lambda_{\max}\left(\Sigma(\tau)^\top P\Sigma(\tau)\right)
\end{aligned}
$$ 
are well-defined strictly positive real numbers. In particular, the existence of the above lower bound
ensures that \eqref{eq:Sand} holds.

Now we show the satisfaction of \eqref{eq:Vdot}. Straightforward calculations yields, for all $\tau\in\R$:
\begin{equation}
\begin{aligned}
\frac{d}{d\tau}\Sigma(\tau)=&-A_{cl}^\top \Sigma^\top(\tau)P\Sigma(\tau)-\Sigma^\top(\tau)PA_{cl}\Sigma(\tau)=\\
&-\He\left(\Sigma^\top(\tau)PA_{cl}\Sigma(\tau)\right)
\end{aligned}
\label{eq:diffSigma}
\end{equation}
where the second equality follows from the identity
$A_{cl}\Sigma(\tau)=\Sigma(\tau)A_{cl}$.
Thus, using \eqref{eq:diffSigma} and the definition of $f$ in \eqref{eq:HybCL}, one gets, for all $x\in\mathcal{C}$
\begin{equation}
\label{eq:flow_in}
\langle \nabla V(x), f(x)\rangle=-\sigma V(x),\quad \forall x\in\mathcal{C}.
\end{equation}
The latter shows that \eqref{eq:Vdot} holds with $\lambda_c=\sigma$.

To conclude the proof, next we show that the satisfaction of \eqref{eq:trace_cond} and \eqref{eq:M_matrix} ensures that \eqref{eq:DV} and \eqref{eq:Vmu} hold with $\mu=1$ and some positive $\lambda_d$. Let $x=(\xi, \tau)\in\mathcal{D}$ and $g\in G(x)$. From the definition of the jump set $\mathcal{D}$ one has that $\tau=T$. Moreover, the definition of $G$  and \eqref{eq:PSI_inc} ensure that there exists $v\in\Psi_\Delta(K\xi)$ such that $g=((G_{cl}+J_{cl}K)\xi+J_{cl}v, 0)$. From Lemma~\ref{lemm:Sector} one has
$$
V(g)\leq V(g)-v^\top S_1 v^\top+\Delta^\top S_1\Delta-2 v^\top S_2(v+K\xi)
$$
Adding to both sides $-V(x)+\varrho (V(x)-1)$ and bearing in mind that $\textcolor{blue}{x=(
\xi, T)}$ one gets:
\begin{equation}
\begin{aligned}
V(g)-V(x)+\varrho(V(x)-1)\leq &\begin{bmatrix}
\xi\\
v
\end{bmatrix}^\top M\begin{bmatrix}
\xi\\
v
\end{bmatrix}+\Delta^\top S_1\Delta-\varrho\\
&+\xi^\top P(1-\exp(-\sigma T))\xi
\end{aligned}
\label{eq:DeltaV_theo}
\end{equation}
with $M$ defined as in \eqref{eq:M_matrix}. Thus, thanks to \eqref{eq:trace_cond} and  \eqref{eq:M_matrix}, from \eqref{eq:DeltaV_theo} one gets:
\begin{equation}
\label{eq:DeltaVgen}
V(g)-V(x)+\varrho(V(x)-1)\leq \xi^\top\underbrace{(-\beta I+P(1-\exp(-\sigma T)))}_{Q}\xi
\end{equation}
with $\beta\coloneqq\vert\lambda_{\max}(M)\vert$. At this stage, let $\sigma^\star>0$ be small enough so that $Q\prec 0$. This selection is always possible. Pick any $\sigma\in (0, \sigma^\star]$, hence, using \eqref{eq:SandOmega}, from \eqref{eq:DeltaVgen} one gets 
$$
V(g)-V(x)+\varrho(V(x)-1)\leq -\alpha \omega_2^{
-1}V(x)
$$
with $\alpha\coloneqq\vert\lambda_{\max}(Q)\vert$.
Thus, if $x\in\mathcal{D}\setminus L_V(1)$ one has
$
V(g)-V(x)\leq -\alpha\omega_2^{
-1}V(x)$, 
which leads to $V(g)\leq e^{-\lambda_d}V(x),$
with $\lambda_d\coloneqq \ln\max\left\{1-\alpha\omega_2^{
-1}, \varsigma\right\},$
where $\varsigma>0$ is an arbitrarily small scalar. Since the analysis holds for any $x\in\mathcal{D}\setminus L_V(1)$, $g\in G(x)$, \eqref{eq:DV} holds with $\mu=1$. 

To conclude the proof, we show that \eqref{eq:Vmu} follows from the satisfaction of  \eqref{eq:trace_cond} and \eqref{eq:M_matrix}. To this end, pick $x\in\mathcal{D}, g\in G(x)$. Then, from \eqref{eq:DeltaVgen} one has:
$$
0\geq V(g)-V(x)+\varrho(V(x)-1)=V(g)-1+(1-\varrho)(1-V(x)).
$$
Thus, since $\varrho\in (0, 1)$ if $x\in\mathcal{D}\cap L_V(1)$, the latter implies
$V(g)\leq 1$, 
which corresponds to  \eqref{eq:Vmu}. This establishes the result.  \QEDA
\end{pf}

Condition  \eqref{eq:M_matrix} is a nonlinear matrix inequality. Handling this type of expressions is generally difficult from a numerical standpoint. Next, we provide an equivalent condition to \eqref{eq:M_matrix} that takes the form of a bilinear matrix inequality (BMI). This equivalent characterization is extensively exploited in Section~\ref{sec:CompDesign}. 
\begin{proposition}
\label{prop:EquivM}
There exist $S_1, S_2\in\mathbb{D}^{n_u}_{++}$, $\varrho\in(0, 1)$, $P\in\mathbb{S}^{n_p+n_s}_{++}$, and $K\in\mathbb{R}^{n_u\times (n_p+n_u)}$ such that 
\eqref{eq:M_matrix} holds if and only if
\begin{equation}
\begin{bmatrix}
(\varrho-1)P&-K^\top S_2&(G_{cl}+J_{cl}K)^\top\Gamma(P)\\
\star&-S_1-2S_2&J_{cl}^\top\Gamma(P)&\\
\star&\star&-\Gamma(P)
\end{bmatrix}\prec 0,
\label{eq:M_MI_2}
\end{equation}
where 
$$
\Gamma(P)\coloneqq \exp(A^\top_{cl}T) P\exp(A_{cl}T).
$$
\end{proposition}
\begin{pf}
Consider the following equivalent rewriting of the matrix $M$ in \eqref{eq:M_matrix}:
$$
\begin{aligned}
M=\begin{bmatrix}
(\varrho-1)P&-K^\top S_2\\
\star&-S_1-2S_2
\end{bmatrix}\\
&\hspace{-2cm}+\begin{bmatrix}
(G_{cl}+J_{cl}K)^\top\\
J_{cl}^\top
\end{bmatrix}\Gamma(P)\begin{bmatrix}
G_{cl}+J_{cl}K&
J_{cl}
\end{bmatrix}
\end{aligned}
$$
Then, from Schur complement lemma $M\prec 0$ and $\Gamma(P)\succ 0$ if and only if
$$
\begin{bmatrix}
(\varrho-1)P&-S_2K&(G_{cl}+J_{cl}K)^\top \\
\star&-S_1-2S_2&J_{cl}^\top\\
\star&\star&-\Gamma(P)^{-1}
\end{bmatrix}\prec 0.
$$
The latter, via a simple congruence transformation, is equivalent to 
\eqref{eq:M_MI_2}. The result is established. \QEDA 
\end{pf}

\section{Controller design}
\label{sec:CompDesign}
In this section, we show how the results developed in Section~\ref{sec:SuffCond} can be exploited to devise a design algorithm for the solution to Problem~\ref{pb:Design} based on semidefinite programming. In particular, the implicit objective of Problem~\ref{pb:Design} is to design a controller gain $K$ 
minimizing the size of the attractor $\mathcal{A}$ defined in \eqref{eq:Attract}. In this regard, it is worth to observe that for the attractor $\mathcal{A}$ in \eqref{eq:Attract} the following chain of inclusions holds:
\begin{equation}
\label{eq:Aincl}
\begin{aligned}
\mathcal{A}\subset&\underbrace{\left(\bigcup_{\tau\in [0, T]}\mathcal{E}\left(\exp(A_{cl}\tau)^\top P\exp(A_{cl}\tau)\right)\right)\times [0, T]}_{\mathcal{O}}\\
 &\subset\mathcal{E}\left(\varpi\lambda_{\min}(P)I\right)\times [0, T]
 \end{aligned}
\end{equation}
with 
$$
\varpi\coloneqq \min_{\tau\in[0, T]}\lambda_{\min}\left(\exp(A_{cl}\tau)^\top\exp(A_{cl}\tau)\right).
$$
This shows that, $\varpi$ being fixed, the attractor $\mathcal{A}$ can be shrunk by maximizing $\lambda_{\min}(P)$. In particular, this naturally leads to formulate Problem~\ref{pb:Design} as the following optimization problem:
\begin{equation}
\label{eq:opti}
\begin{aligned}
& \underset{P, K, S_1, S_2, \varrho, c}{\maximize}
& & c\\
& \text{subject to}
& & P-c I\succeq 0, \eqref{eq:trace_cond}-\eqref{eq:M_matrix}\\
&&& \varrho\geq 0, P\in\mathbb{S}^{n_p+n_c}_{++}, S_1, S_2\in\mathbb{D}^{n_u}_{++}\\
&&&K\in\mathbb{R}^{n_u\times (n_p+n_u)}.
\end{aligned}
\end{equation}
It is important to observe that due to \eqref{eq:M_MI_2} being bilinear in the decision variables,  \eqref{eq:opti} is numerically intractable. To overcome this problem, in this paper we rely on the approach proposed in \cite{dinh2011combining} to suboptimally solve \eqref{eq:opti}
via a sequence of semidefinite programming (\emph{SDP}) problems, i.e., optimization problems with linear objective over linear matrix inequality constraints. To deploy this approach,  as a first step we rewrite \eqref{eq:M_MI_2} in the following equivalent linear-bilinear decomposed form:
$$
\mathcal{L}(P, S_1, S_2)+\He(\mathcal{X}^\top(\varrho, K)\mathcal{Y}(P, S_2))\prec 0,
$$
with 
$$
\begin{aligned}
&\mathcal{L}(P, S_1, S_2)\coloneqq\begin{bmatrix}
      -P & 0 & G_{cl}^\top\Gamma(P)\\[0.3em]
       \star & -S_1-2S_2& J_{cl}^\top \Gamma(P)\\[0.3em]
       \star & \star & -\Gamma(P)\\[0.3em]
     \end{bmatrix},\\ 
     &\mathcal{X}^\top(\varrho, K)\coloneqq\begin{bmatrix}
\frac{\varrho}{2}I&K^\top\\[0.3em]
0&0\\[0.3em]
0&0
 \end{bmatrix}, \mathcal{Y}(P, S_2)\coloneqq\begin{bmatrix}
P&0&0\\[0.3em]
0&-S_2&J_{cl}^\top \Gamma(P)
 \end{bmatrix}.
 \end{aligned}
$$
The latter, dropping the dependency on the decision variables, can be equivalently rewritten in the following psd convex-concave decomposed form; see  \cite[Definition 2.1]{dinh2011combining}:
$$
\mathcal{L}+\mathcal{X}^\top\mathcal{X}+\mathcal{Y}^\top\mathcal{Y}-(\mathcal{X}-\mathcal{Y})^\top (\mathcal{X}-\mathcal{Y})\prec 0
$$
which, by Schur complement's lemma, is equivalent to:
\begin{equation}
\begin{bmatrix}
\mathcal{L}-\mathcal{X}^\top\mathcal{X}-\mathcal{Y}^\top\mathcal{Y}+\He(\mathcal{X}^\top\mathcal{Y})&\mathcal{X}^\top&\mathcal{Y}^\top\\
\star&-I&0\\
\star&\star&-I
\end{bmatrix}\prec 0.
\label{eq:ConvConc}
\end{equation}
Based on the equivalent rewriting of \eqref{eq:M_MI_2} in \eqref{eq:ConvConc}, the following ``linear inner approximation''  of optimization problem \eqref{eq:opti} around the point $\eta^{(0)}=(\varrho^{(0)}, P^{(0)}, K^{(0)}, S_2^{(0)})\in\R\times\mathbb{S}_{++}^{n_p+n_u}\times\R^{n_u\times (n_p+n_u)}\times\mathbb{D}^{n_u}_{++}$ is considered:
$$
O^{\eta^{(0)}}\colon\left\{\begin{aligned}
& \underset{P, K, S_1, S_2, \varrho, c}{\maximize}
& & c\\
&\text{s.t.}
& &\!\!\!\mathcal{M}(P, S_1, S_2, K, \varrho\vert \eta^{(0)})\prec 0,\\
&&&\!\!\! \Delta^\top S_1\Delta-\varrho\leq 0\\
&&&\!\!\! P-cI\succeq 0, \varrho\in (0, 1)\\
&&&\!\!\! P\in\mathbb{S}^{n_p+n_u}_{++}, S_1, S_2\in\mathbb{D}^{n_u}_{++},
\end{aligned}\right.
$$
where:
$$
\begin{aligned}
\underbrace{\begin{bmatrix}
\mathcal{R}(P, S_1, S_2, K, \varrho\vert \eta^{(0)})&\mathcal{W}(\varrho, P, S_2, K)\\
\star&-I
\end{bmatrix}}_{\mathcal{M}(P, S_1, S_2, K, \varrho\vert \eta^{(0)})},
\end{aligned}
$$
$$
\begin{aligned}
&\mathcal{R}(P, S_1, S_2, K,  \varrho\vert \eta^{(0)})\coloneqq\mathcal{L}(P, S_1, S_2, P)+\mathcal{Q}(\eta^{(0)})\\
&+(D\mathcal{Q}(\eta^{(0)}))(\varrho-\varrho^{(0)}, P-P^{(0)}, K-K^{(0)}, S_2-S_2^{(0)})\\
&\mathcal{W}(\varrho, P, S_2, K)\coloneqq\begin{bmatrix}\mathcal{X}^\top(\varrho, K)&\mathcal{Y}^\top(P, S_2)\end{bmatrix},
\end{aligned}
$$
in which 
$$
\begin{aligned}
(\varrho, P, K, S_2)\mapsto\mathcal{Q}(\varrho, P, S_2, K)\coloneqq&-\mathcal{X}^\top\mathcal{X}-\mathcal{Y}^\top\mathcal{Y}\\
&+\He(\mathcal{X}^\top\mathcal{Y}).
\end{aligned}
$$
At this stage, provided that an initial feasible solution $\eta^{(0)}$ to optimization problem~\eqref{eq:opti} is available, $O^{\eta^{(0)}}$ is solved. This, due to the property of psd-convex functions; see \cite[Lemma 2.1, item (b)]{dinh2011combining}, generates another feasible solution to optimization problem~\eqref{eq:opti}. Then, the process is repeated, thereby giving rise to a sequence of SDP problems.  
The applicability of this approach requires the knowledge of an initial feasible solution to optimization problem~\eqref{eq:opti}, which is basically used to generate the sequence of SDP problems $\{O^{\eta^{(k)}}\}$.  To this end, we make use of the following result. 
\begin{proposition}
Suppose that there exist $W\in\mathbb{S}^{n_p+n_u}_{++}$, $S_1\in\mathbb{D}_{++}^{n_u}$, $Y\in\R^{n_u\times(n_p+n_u)}$, and $\varrho\in (0, 1)$ such that 
\eqref{eq:trace_cond} holds and
\begin{equation}
\label{eq:M0}
\begin{bmatrix}
(\varrho-1)W&0&W G_{cl}^\top +Y^\top J_{cl}^\top\\
\star&-S_1&J_{cl}^\top&\\
\star&\star&-\Theta(W)
\end{bmatrix}\prec 0
\end{equation}
where:
$$
\Theta(W)\coloneqq \exp(-A_{cl}T)W\exp(-A_{cl}^\top T).
$$
Then, there exists $S_2\in\mathbb{D}_{++}^{n_u}$ such that  $(\varrho, S_1, S_2, P=W^{-1}, K=W^{-1}Y)$ is a feasible solution to \eqref{eq:trace_cond}-\eqref{eq:M_MI_2}. Moreover, 
let the pair
\begin{equation}
\label{eq:pairAB_D}
\begin{aligned}
&A_p^D\coloneqq\exp(A_p T), &B_p^D\coloneqq\int_0^T \exp(A_ps)B_pds.
\end{aligned}
\end{equation}
be discrete-time stabilizable. Then, \eqref{eq:trace_cond} and \eqref{eq:M0} are feasible. 
\label{lemm:LMI}
\end{proposition}
\begin{pf}
A simple congruence transformation shows that the satisfaction of \eqref{eq:M0} is equivalent to:
\begin{equation}
\label{eq:M0_equiv_S20}
\begin{bmatrix}
(\varrho-1)P&0&(G_{cl}+J_{cl}K)^\top\Gamma(P)\\
\star&-S_1&J_{cl}^\top\Gamma(P)&\\
\star&\star&-\Gamma(P)
\end{bmatrix}\prec 0
\end{equation}
which corresponds to \eqref{eq:M_MI_2} with $S_2=0$.
Hence, standard perturbation arguments on the previous strict inequality ensure the existence of $\epsilon>0$ small enough such that for any $S_2\in\mathbb{D}_{++}^{n_u}$ with $\Vert S_2\Vert\leq \epsilon$, $(\varrho, S_1, S_2, P=W^{-1}, K=W^{-1}Y)$ fulfills \eqref{eq:trace_cond}-\eqref{eq:M_MI_2}. This concludes the first part of the proof. To conclude the proof, observe that from Proposition~\ref{prop:feasible} in the Appendix, under the discrete-time stabilizability of the pair $(A_p^D, B_p^D)$, there exists a feasible solution to \eqref{eq:trace_cond}-\eqref{eq:M0_equiv_S20}.  Hence, bearing in mind the equivalence between \eqref{eq:M0}  and \eqref{eq:M0_equiv_S20} established above, the proof is concluded.
\hfill \QEDA
\end{pf}

In light of Proposition~\ref{lemm:LMI}, to select a feasible initial point to optimization problem~\eqref{eq:opti}
we proceed as follows. First, we determine a solution to the following optimization problem:
\begin{equation}
\label{eq:opti2}
\begin{aligned}
& \underset{W, S_1, \varrho, Y}{\minimize}
& & l\\
& \text{subject to}
&&\eqref{eq:M0}\\
   & & &\Delta^\top S_1\Delta-\varrho\leq 0\\
   & &  &W-l I\preceq 0, \varrho\in(0, 1)\\
   &&& W\in\mathbb{S}^{n_p+n_u}_{++}, S_1\in\mathbb{D}^{n_u}_{++}, Y\in\mathbb{R}^{n_u\times (n_p+n_u)}.
\end{aligned}
\end{equation}

Proposition~\ref{lemm:LMI} ensures that \eqref{eq:opti2} is always feasible provided that the pair 
$(A_p^D, B_p^D)$ defined in \eqref{eq:pairAB_D} is discrete-time stabilizable. Once a feasible solution to  \eqref{eq:opti2} is available, a feasible solution $(W, S_1, S_2, \varrho, Y)$ to optimization problem~\eqref{eq:opti} can be easily obtained by selecting $P$ and $K$ as  indicated in Proposition~\ref{lemm:LMI} and $S_2\in\mathbb{D}^{n_u}_{++}$ so to fulfill \eqref{eq:M_MI_2}. 

It is important to remark that whenever $\varrho$ is fixed, \eqref{eq:opti2} is an SDP problem. Thus a feasible solution to \eqref{eq:opti2} can be easily computed by performing a line search on the variable $\varrho$ in the interval $(0, 1)$. 

Based on the steps presented so far and on the equivalence between \eqref{eq:M_matrix} and \eqref{eq:M0} established in Proposition~\ref{lemm:LMI}, our approach to solve optimization problem~\eqref{eq:opti} is summarized in Algorithm~\ref{alg:design}.\IncMargin{2em}
\begin{algorithm}
\caption{Optimal controller design}
\label{alg:design}
\Indm
\KwIn{Matrices $A_p, B_p$, quantization levels vector $\Delta\in\R^{n_u}_{>0}$, sampling-time $T>0$, $k_{\max}\in\mathbb{N}_{>0}$, and $\varepsilon>0$.}
\Indp
  \BlankLine
  
\textbf{Initial solution:} Solve \eqref{eq:opti2} via a line search on $\varrho\in(0, 1)$. 
Let $W^{(0)}$, $\varrho^{(0)}$, $S_1^{(0)}$ and $Y^{(0)}$ be the values associated to the corresponding solution.  
\medskip

Set $k=0$ and $P^{(0)}=(W^{(0)})^{-1}, K^{(0)}=P^{(0)}Y^{(0)}$\;
Fix $(P, S_1, \varrho, K)=(P^{(0)}, S_1^{(0)}, \varrho^{(0)}, K^{(0)})$ and solve LMI \eqref{eq:M_MI_2} with respect to $S_2$. Let $S_2^{(0)}$ be the corresponding solution\;
\While{$k<k_{\max}$}{
Solve SDP problem $O^{(\varrho^{(k)}, P^{(0)}, K^{(0)}, S_2^{(0)})}$\;
$\varrho^{(k+1)}\longleftarrow \varrho$, $P^{(k+1)}\longleftarrow P$, $K^{(k+1)}\longleftarrow K$, $S_2^{(k+1)}\longleftarrow S_2$\;
\SetAlgoNoLine
\If{$\vert c^{(k+1)}-c^{(k)}\vert\leq \varepsilon$}{break\;}
$k\longleftarrow k+1$\;
}
\KwRet{$K$ and $P$}
\end{algorithm}
\section{Numerical Simulations}
\label{sec:Ex}
\begin{figure}
\centering
\psfrag{t}[1][1][1]{$t$}
\psfrag{x1}[1][1][1]{$x_{p1}$}
\psfrag{x2}[1][1][1]{$x_{p2}$}
\includegraphics[trim=0.1cm 0 0.1cm 0.5cm, clip, width=1\columnwidth]{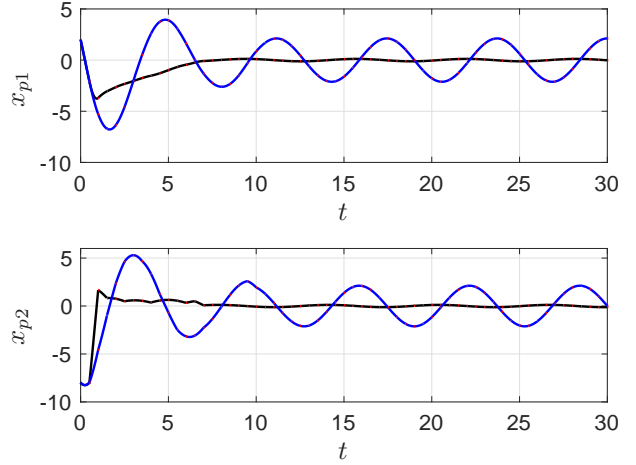}
\caption{\label{fig:sims} Evolution of the plant state in closed loop with: optimal gain (black line) and LQR gain (blue line).}
\end{figure}

\begin{figure}
\centering
\psfrag{chi}[1][1][1]{$\chi$}
\psfrag{x1}[1][1][1]{$x_{p1}$}
\psfrag{x2}[1][1][1]{$x_{p2}$}
\includegraphics[trim=0.1cm 0 0.1cm 0.5cm, clip, width=1\columnwidth]{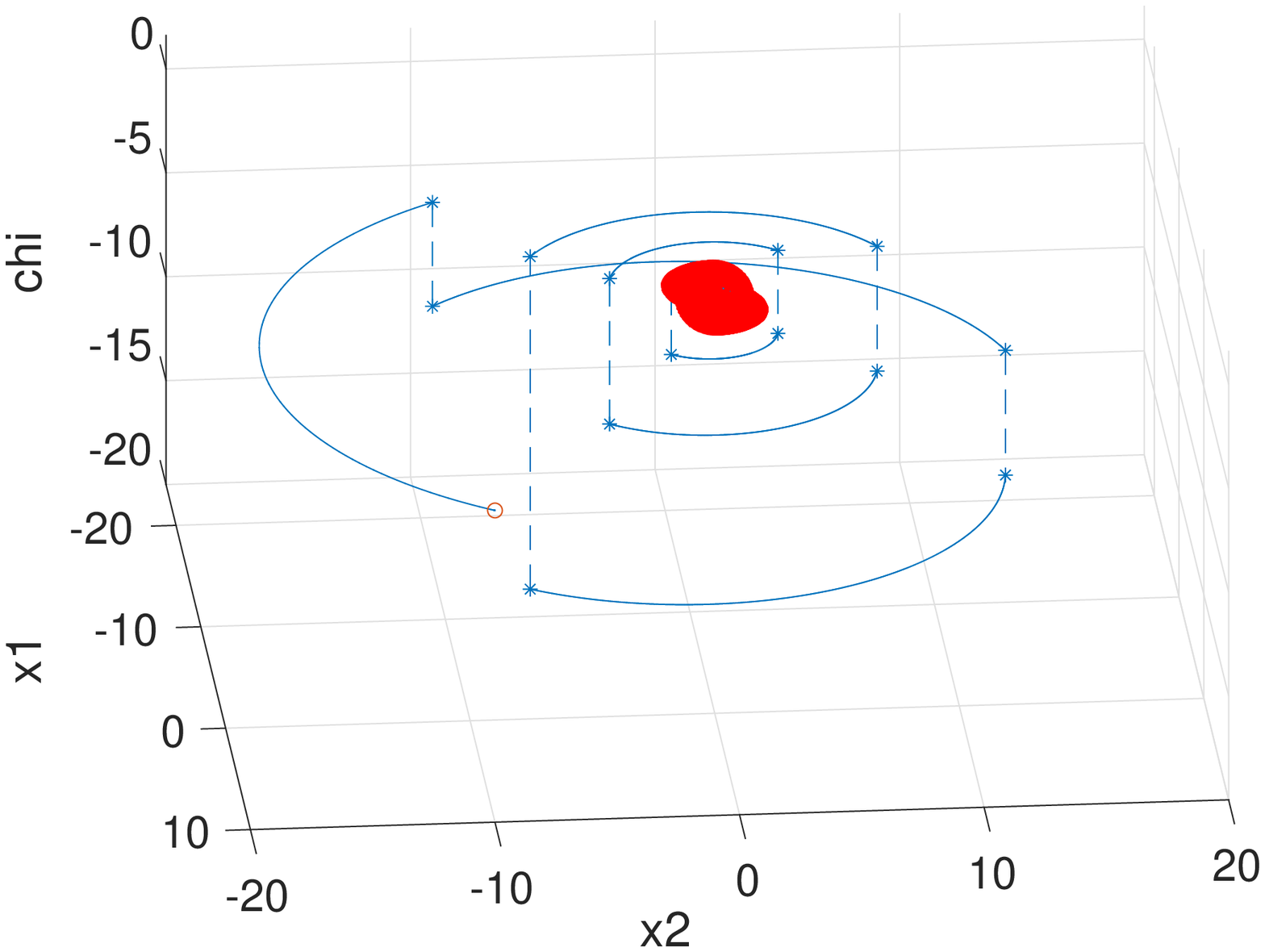}\\
\includegraphics[trim=0.1cm 0 0.1cm 0.5cm, clip, width=1\columnwidth]{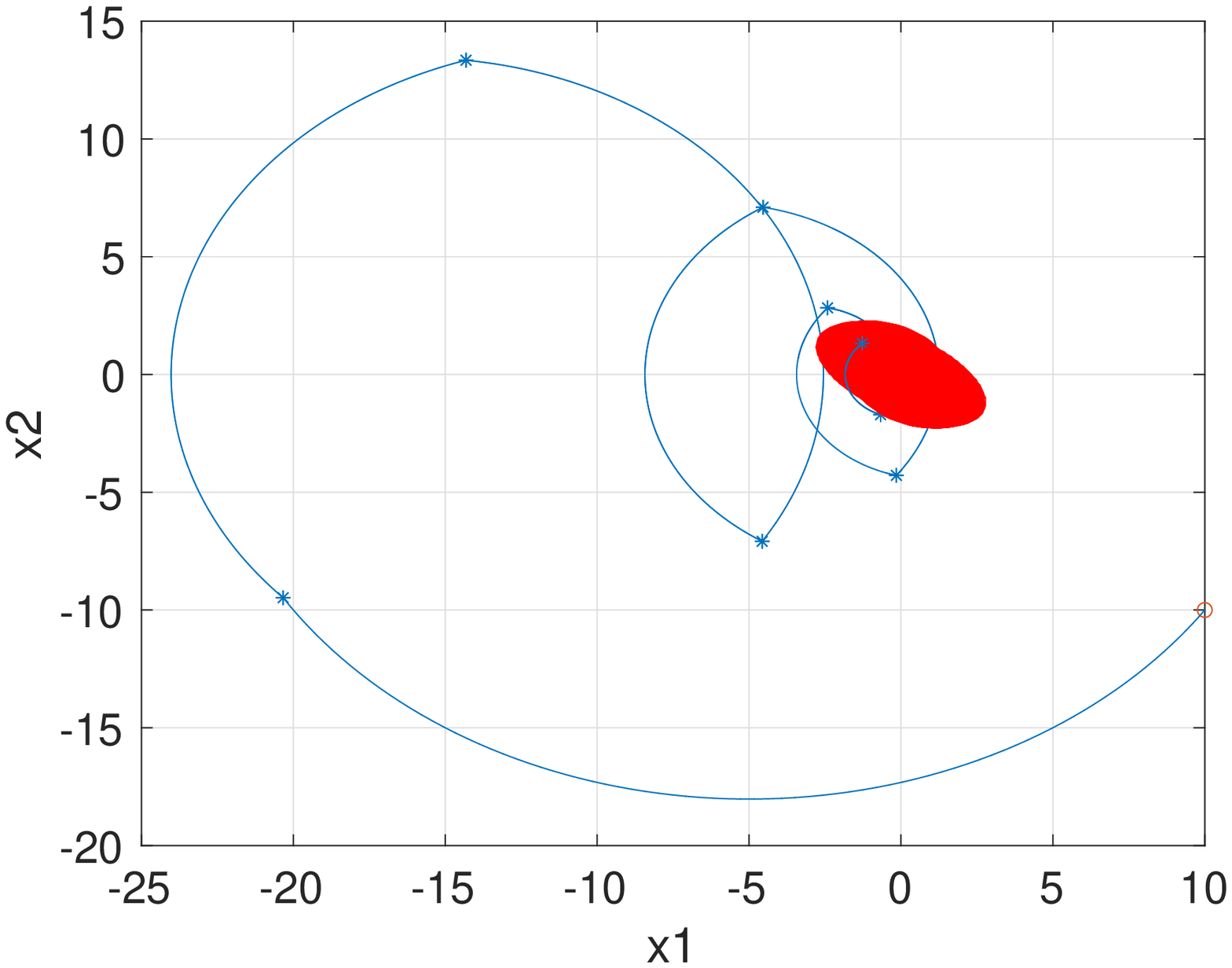}
\caption{\label{fig:portrait} Top figure: Evolution of the closed-loop state $\xi$ from $x(0, 0)=(10, -10, -5, 0)$ (blue line) and projection of the outer estimate of the attractor $\mathcal{A}$ (red line). Bottom figure: A view from the ``top''.}
\end{figure}

We consider the following data for system \eqref{1}:
$$
A_p=\begin{bmatrix}0 &1\\ -1&0\end{bmatrix}, B_p=\begin{bmatrix}0\\1\end{bmatrix}, \Delta=1
$$
and select $T=0.5$. Setting $\varepsilon=10^{-4}$, Algorithm~\ref{alg:design} terminates in 75 iterations and
yields\footnote{Numerical solutions to LMIs are obtained through the solver \textit{SDPT3} \cite{tutuncu2003solvingSDPT3} and coded in MATLAB$^{\tiny{\textregistered}}$ via \textit{YALMIP} \cite{lofberg2004yalmip}. Simulations of hybrid dynamical systems are performed in MATLAB$^{\tiny{\textregistered}}$ via the \emph{Hybrid Equations Toolbox} \cite{sanfelice2013toolbox}.}: 
$$\begin{array}{lcl}
K&=&\begin{bmatrix}
0.5529&   -2.1873  & 0
\end{bmatrix},\\
P&=&\begin{bmatrix}
2.5432 &   0.0046   & 0.0009\\
    0.0046   & 2.5432  & -0.0007\\
    0.0009  & -0.0007    &0.4916
\end{bmatrix}.
\end{array}
$$ 
To show the effectiveness of Algorithm~\ref{alg:design} in designing a controller inducing a ``small'' attractor, 
in \figurename~\ref{fig:sims} we report numerical simulation of the plant state in closed loop with 
the gain $K$ returned by Algorithm~\ref{alg:design} and with a gain designed via LQR on the discretized version of the plant, with a generic (non necessarily optimal) tuning of the weighting matrices $Q$ and $R$. \figurename~\ref{fig:sims} conveys two important messages. First, it clearly shows that the selection of the control gain plays a relevant role on the response the closed-loop system. Second, it emphasizes the effectiveness of the proposed approach towards the objective of confining the plant state close to the origin. \figurename~\ref{fig:portrait} depicts the evolution of the $\xi$-component of the closed-loop state along with the projection of the outer estimate $\mathcal{O}$ of the attractor $\mathcal{A}$ defined in \eqref{eq:Aincl}.
\section{Conclusion}
The problem of designing a sampled-data state feedback control law for continuous-time linear control systems subject to uniform input quantization was addressed. The main goal was to design a sampled-data state feedback to ensure uniform global asymptotic stability (UGAS) of a compact set containing the origin of the plant substate. The closed loop was augmented with an auxiliary clock variable triggering the occurrence of sampling events and modeled as a hybrid dynamical system. Sufficient conditions for stability in the form of bilinear matrix inequalities were proposed. An iterative algorithm based on the use of the convex-concave decomposition approach was proposed to design an optimal controller with the objective of minimizing the size of the closed-loop attractor. Under some mild conditions on the open-loop system, the proposed design algorithm was shown to be deadlock free. The interest of the proposed methodology was illustrated through a numerical example.

This work paves the ways for future research on quantized sampled-data control design. Extensions to aperiodic sampled-data systems as well as to other classes of quantizers are currently part of our research. 
\bibliography{biblio}
\appendix 
\section{Ancillary results and definitions}
\label{app:results}
\subsection{Preliminaries on Hybrid Systems}
In this paper, we consider hybrid systems with state $x\in\R^n$ of the form: 
\begin{equation}
\label{eq:Hybrid}
\mathcal{H}\begin{cases}
\dot{x}=f(x),& x\in\mathcal{C}\\
x^+\in G(x), & x\in\mathcal{D}.
\end{cases}
\end{equation}
where $f\colon\R^n\to\R^n$ is the  \emph{flow map}, $\mathcal{C}$ is the \emph{flow set}, $G\colon\R^n\rightrightarrows\R^n$ is the set-valued \emph{jump map}, and $\mathcal{D}$ is the \emph{jump set}. A set $E\subset\R_{\geq 0}\times \mathbb{N}$ is a \emph{hybrid time domain} if it is the union of a finite or infinite sequence of intervals $[t_j, t_{j+1}]\times\{j\}$, with the last interval (if existent) of the form $[t_j,T)$ with $T$ finite or $T=\infty$. Given a hybrid time domain $E$, we denote by $\sup_t E=\sup \{t\in\mathbb{R}_{\geq 0}\colon \exists j\in\mathbb{N}\,\mbox{s.t.}\,(t, j)\in E\}$. A function $\phi\colon \dom\phi\rightarrow\R^{n}$ is a hybrid arc if $\dom\phi$ is a hybrid-time domain and $\phi(\cdot, j)$ is locally absolutely continuous for each $j$. 
A hybrid arc $\phi$ is a solution to \eqref{eq:Hybrid} if it satisfies the dynamics of \eqref{eq:Hybrid}. A solution $\phi$ to \eqref{eq:Hybrid} is maximal if it cannot be extended and is complete if $\dom\phi$ is unbounded. In particular, given $S\subset\R^{n}$, we denote by $\mathcal{S}_{\mathcal{H}}(S)$ the set of maximal solutions $\phi $ to $\mathcal{H}$ with $\phi(0,0)\in S$; see~\cite{goebel2012hybrid} for more details on solutions to hybrid systems. 

The following notion of asymptotic stability for compact sets is considered in this paper. 
\begin{definition}
Let $\mathcal{A}\subset\R^n$ be compact. We say that $\mathcal{A}$ is: 
\begin{itemize}
\item stable for $\mathcal{H}$, if for all $\varepsilon>0$, there exists $\delta>0$ such that any $\phi\in\mathcal{S}(\mathcal{A}+\delta\mathbb{B})$ is such that
$$
\vert\phi(t, j)\vert_{\mathcal{A}}\leq \varepsilon\quad \forall (t, j)\in\dom\phi;
$$
\item uniformly globally attractive (UGA) for $\mathcal{H}$, if all maximal solutions to $\mathcal{H}$ are complete and for any $\mu>0, \varepsilon>0$, there exists $T>0$ such that for all $\phi\in\mathcal{S}(\mathcal{A}+\mu\mathbb{B})$ the following implication holds:
$$
(t, j)\in\dom \phi, t+j\geq T\implies \vert \phi(t, j)\vert_{\mathcal{A}}\leq \varepsilon;
$$
\item uniformly globally asymptotically stable (UGAS) for $\mathcal{H}$, if it is both stable and uniformly globally attractive.
\end{itemize}
\end{definition}
\begin{definition}
Let $S\subset\R^n$, we say that $S$ is strongly forward invariant for $\mathcal{H}$ if any $\phi\in\mathcal{S}(S)$ is complete and such that $\rge\phi\subset S$.
\end{definition}
%
\begin{proposition}
\label{prop:gen_stab}
Let $\mathcal{H}$ satisfy the hybrid basic conditions. Suppose that there exist $V\colon\R^{n}\to\R_{\geq 0}$ continuously differentiable on an open 
neighborhood of $\mathcal{C}$, and $\mu, \lambda_d, \lambda_c>0$ such that:
\begin{subequations}
\begin{align}
\label{eq:Sand}
&\lim_{\substack{x\to\infty\\x\in\mathcal{C}\cup\mathcal{D}}}V(x)=\infty\\
\label{eq:Vdot}
&\langle \nabla V(x), f\rangle \leq -\lambda_c V(x)&\forall x\in \mathcal{C}\setminus L_V(\mu), f\in F(x)\\
\label{eq:DV}
&V(g) \leq \exp(-\lambda_d) V(x) &\forall x\in\mathcal{D}\setminus L_V(\mu), g\in G(x)\\
\label{eq:Vmu}
&V(g) \leq \mu&\forall x\in\mathcal{D}\cap L_V(\mu), g\in G(x)
\end{align}
\label{eq:LyapunovResult}
\end{subequations}

Assume that $L_V(\mu)$ is nonempty. Then, if any maximal solution to $\mathcal{H}$ is complete, $L_V(\mu)$ is globally asymptotically stable for $\mathcal{H}$.
\end{proposition}
\begin{pf}
Now, observe that \eqref{eq:Vdot}, \eqref{eq:DV}, and \eqref{eq:Vmu} ensure 
that for any $\rho\geq\mu$, the set $L_V(\rho)$ is strongly forward invariant for $\mathcal{H}$. Hence, since, thanks to \eqref{eq:Sand} sublevel sets of $\left.V\right\vert_{\mathcal{C}\cup\mathcal{D}}$ are bounded, it turns out that any solution to $\mathcal{H}$ is bounded. To conclude the proof, we rely on the following claim, whose proof is reported next.
\begin{claim}
\label{claim:FiniteTime}
Consider hybrid system $\mathcal{H}$ in \eqref{eq:Hybrid}. Suppose that there exist $V\colon \R^{n}\to\R_{\geq 0}$ continuously differentiable on an open 
neighborhood of $\mathcal{C}$ and $\mu, \lambda_d, \lambda_c>0$ 
such that all the conditions in Proposition~\ref{prop:gen_stab} hold. Define for all $\xi\in\R^n$
\begin{equation}
\Upsilon(\xi)\coloneqq\begin{cases}
\frac{1}{\gamma}\ln\left(\frac{V(\xi)}{\mu}\right)&\text{if}\,\,V(\xi)>\mu\\
0&\text{else}.
\end{cases}
\label{eq:Upsilon}
\end{equation}
Let $\phi$ be a complete solution to $\mathcal{H}$. Then, the following holds: 
$$
(t, j)\in\dom\phi,  t+j\geq \Upsilon(\phi(0,0))\implies V(\phi(t, j))\leq \mu.
$$
\hfill$\diamond$
\end{claim}
Hinging upon Claim~\ref{claim:FiniteTime}, next we show that $L_V(\mu)$ is globally uniformly attractive for $\mathcal{H}$. Let $\Upsilon$ be defined as in \eqref{eq:Upsilon}. Observe that $\Upsilon$ is locally bounded on $\R^n$, in fact continuous. Now, pick $r>0$ and let
$$
T_r=\sup_{\xi\in L_V(\mu)+r\mathbb{B}}\Upsilon(\xi).
$$
Notice that, due to $\Upsilon$ being continuous on $\R^n$ and $L_V(\mu)$ being compact and nonempty, $T_r$ is finite and strictly positive. Then, by the definition of $\Upsilon$, it follows that for all $\phi\in\mathcal{S}(L_V(\mu)+r\mathbb{B})$, any $(t, j)\in\dom\phi$ with $t+j\geq T_r$ implies that $\phi(t, j)\in L_V(\mu)$. This implies that $L_V(\mu)$ is UGA for $\mathcal{H}$. Bearing in mind that $\mathcal{H}$ satisfies the hybrid basic conditions and that $L_V(\mu)$ is strongly forward invariant and UGA for $\mathcal{H}$, by invoking \cite[Proposition 7.5, page 142]{goebel2012hybrid}, it follows that $L_V(\mu)$ is stable. This concludes the proof.  \QEDA
\end{pf}
{\bf Proof of Claim~\ref{claim:FiniteTime}.}
Under the considered assumptions, it follows that solutions to $\mathcal{H}$ cannot leave $L_V(\mu)$. Hence, the statement of the result holds for complete solutions to $\mathcal{H}$ from $L_V(\mu)$. Next the show that the assert holds true for all complete solutions. 
By contradiction, assume that there exists a complete solution $\phi$ to $\mathcal{H}$ with $\phi(0, 0)\notin L_V(\mu)$ such that $(t, j)\in\dom\phi$, $t+j\geq \Upsilon(\phi(0, 0))$ imply $V(\phi(t, j))>\mu$. Since solutions to $\mathcal{H}$ cannot escape $L_V(\mu)$, one has that $\rge\phi\cap L_V(\mu)=\emptyset$. Then, by using \eqref{eq:Vdot} and \eqref{eq:DV}, direct integration of $(t, j)\mapsto (V\circ\phi)(t, j)$ yields
$$
V(\phi(t, j))\leq e^{-(\lambda_c t+\lambda_d j)}V(\phi(0, 0))\quad \forall (t, j)\in\dom\phi
$$
which implies
\begin{equation}
\label{eq:Vphi}
V(\phi(t, j))\leq e^{-\gamma(t+j)}V(\phi(0, 0))\quad \forall (t, j)\in\dom\phi.
\end{equation}
Pick $(t, j)\in\dom\phi$ with $ t+j\geq \Upsilon(\phi(0, 0))$. Notice that such a pair $(t, j)$ exists due to $\phi$ being complete and $\Upsilon(\phi(0, 0))>0$. Then, from \eqref{eq:Vphi} one has
$$
V(\phi(t, j))\leq \mu.
$$ 
This contradicts that $\rge\phi\cap L_V(\mu)=\emptyset$. Hence, the assert is proven.  \QEDA
\subsection{Preliminaries on matrix-valued functions}
In this paper, we consider matrix valued functions of the form:
\begin{equation}
\label{eq:functionX}
X\colon \mathcal{S}\to\mathcal{Y}
\end{equation}
where $\mathcal{S}$ is a finite dimensional real linear vector space and $\mathcal{Y}\subset\R^{n\times m}$. 
\medskip

\begin{definition}(Differential)
Let $X$ be defined as in \eqref{eq:functionX}. We say that $X$ is differentiable at $x\in\mathcal{S}$ if there exists a linear map $DX(x)\colon\mathcal{S}\to\mathcal{Y}$ such that:
$$
\lim_{\Vert h\Vert_{\mathcal{S}}\to 0}\frac{\Vert X(x+h)-X(x)-DX(x)h\Vert_{\mathcal{Y}}}{\Vert h\Vert_{\mathcal{S}}}=0
$$
where $\Vert\cdot\Vert_{\mathcal{S}}$ and $\Vert\cdot\Vert_{\mathcal{Y}}$ are any norms, respectively, on $\mathcal{S}$ and $\mathcal{Y}$.\hfill$\diamond$
\end{definition}
\begin{definition}[\cite{shapiro1997first}]
Let $\mathcal{C}\subset\mathcal{S}$ be convex and $\mathbb{S}^n$ be the set of $n\times n$ symmetric matrices. A function $X\colon \mathcal{C}\to\mathbb{S}^n$ is said to be positive semidefinite convex (\emph{psd-convex}) on $\mathcal{C}$ if for all $x, y\in\mathcal{C}$ and $t\in [0, 1]$ the following holds:
$$
X(t x+(1-t)y)\preceq tX(x)+(1-t)X(y).
$$
Furthermore, we say that $X$ is positive semidefinite concave (\emph{psd-concave}) if $-X$ is psd-convex.\hfill$\diamond$
\end{definition}
\begin{lemma}(\cite{dinh2011combining})
\label{lemm:DiffConv}
Let $\mathcal{C}\subset\mathcal{S}$ be convex, $\mathbb{S}^n$ be the set of $n\times n$ symmetric matrices, and $X\colon \mathcal{C}\to\mathbb{S}^n$ be differentiable on an open neighborhood of $\mathcal{C}$. Then, $X$ is psd-convex on $\mathcal{C}$ if and only if for all $x, y\in\mathcal{C}$
$$
X(y)-X(x)\succeq DX(x)(y-x). 
$$\hfill$\diamond$
\end{lemma}
\subsection{Technical results}
\begin{proposition}
\label{prop:feasible}
If the pair $(A_p^D, B_p^D)$ defined in \eqref{eq:pairAB_D}  is discrete-time stabilizable, 
then the conditions of Theorem~\ref{theo:main} are feasible. Moreover, for any $\epsilon\geq 0$, there exists a solution to \eqref{eq:trace_cond}-\eqref{eq:M_matrix} with $\Vert S_2\Vert\leq \epsilon$.
\end{proposition}
\begin{pf}
To show the result, we prove that \eqref{eq:trace_cond} and \eqref{eq:M_matrix} (which is equivalent to \eqref{eq:M_MI_2} by the virtue of Proposition~\ref{prop:EquivM}) are feasible with $S_2=0$ and some $P\in\mathbb{S}^{n_p+n_u}_{++}$, $K$, $\varrho\in(0, 1)$, and $S_1\in\mathbb{D}^{n_p+n_u}_{++}$. Notice that, the existence of such a tuple $(P, S_1,\varrho, K)$, by continuity of the matrix $M$ in \eqref{eq:M_matrix} with respect to the entries of $S_2$ and the inequality in \eqref{eq:M_matrix} being strict, ensures, for any $\epsilon\geq 0$, the existence of $\epsilon^\star\in [0, \epsilon]$ such that for any $S_2\in\mathbb{D}^{n_u}_{++}$ with $\Vert S_2\Vert\leq \epsilon^\star$, $(P, S_1, S_2, \varrho, K)$ fulfills \eqref{eq:M_matrix}, thereby establishing the result. 
To this end,  pick $K$ such that $\specr(\exp(A_{cl}T)(G_{cl}+J_{cl}K))<1$. Notice that since 
$$
E\coloneqq\exp(A_{cl}T)(G_{cl}+J_{cl}K)=\begin{bmatrix}
A_p^D&0\\
0&0
\end{bmatrix}+\begin{bmatrix}
B_p^D\\
I
\end{bmatrix}K
$$
under the stabilizability of $(A_p^D, B_p^D)$ the above selection of $K$ is feasible. Pick $\varrho\in (0, 1)$ such that 
     $$
     \sqrt{1-\varrho}>\specr(E),
     $$
     this is always possible due to $\specr(E)<1$. For this selection of $\varrho$, select $S_1\in\mathbb{D}^{n_u}_{++}$ such that \eqref{eq:trace_cond} holds. Let $Q\in\mathbb{S}^{n_p+n_u}_{++}$ be any solution to the following matrix inequality:
\begin{equation}
\label{eq:LMI_DT}
\begin{aligned}
&E^\top QE+(\varrho-1)Q\prec 0.
\end{aligned}
\end{equation}
which is solvable due to the selection of $\varrho$ above. The latter, thanks to Schur complement lemma and a simple congruence transformation, is equivalent to:
     \begin{equation}
 \label{eq:Elimin}
\begin{bmatrix}
       (\varrho-1)Q & E^\top Q\\[0.3em]
       \star & -Q\\[0.3em]
     \end{bmatrix}\prec 0.
\end{equation}
At this stage, observe that \eqref{eq:Elimin} can be equivalently rewritten as:
     \begin{equation}
 \label{eq:Elimin2}
\begin{bmatrix}
I&0&0\\[0.3em]
0&0&I
     \end{bmatrix}
\begin{bmatrix}
       (\varrho-1)Q & 0& E^\top Q\\[0.3em]
       \star & 0& F^\top Q\\[0.3em]
       \star & \star & -Q\\[0.3em]
     \end{bmatrix}\begin{bmatrix}
I&0\\[0.3em]
0&0\\[0.3em]
0&I
\end{bmatrix}\prec 0,
\end{equation}
with $F\coloneqq \exp(A_{cl}T)J_{cl}$. Thus, from the \emph{Projection Lemma} \cite[Lemma 3.1]{gahinet1994linear}, the satisfaction of \eqref{eq:Elimin2} implies that there exists $X\in\R^{n_u\times n_u}$ such that
\begin{equation}
\label{eq:Elim2}
\begin{bmatrix}
       (\varrho-1)Q & 0& E^\top Q\\[0.3em]
       \star & \He(X)& F^\top Q\\[0.3em]
       \star & \star & -Q\\[0.3em]
    \end{bmatrix}\prec 0.
 \end{equation}
 Notice that, necessarily, $\He(X)\prec 0$. Let 
 $$
 \varrho\leq\frac{\lambda_{\min}(S_1)}{\lambda_{\max}(-\He(X))},
 $$
observe that $\varrho>0$, due to $S_1\succ 0$ and $\He(X)\prec 0$. Therefore, by setting 
$$
P\coloneqq \varrho Q, \quad Z\coloneqq \varrho X,
$$
from \eqref{eq:Elim2}, one has
\begin{equation}
\label{eq:Elim3}
\begin{bmatrix}
       (\varrho-1)P & 0& E^\top P\\[0.3em]
       \star  & \He(Z)& F^\top P\\[0.3em]
       \star  &\star  & -P\\[0.3em]
     \end{bmatrix}\prec 0.
\end{equation}
At this stage, notice that by construction, $-S_1-\He(Z)\preceq~0$, therefore \eqref{eq:Elim3} implies
$$
\begin{bmatrix}
       (\varrho-1)P & 0& E^\top P\\[0.3em]
       \star &-S_1& F^\top P\\[0.3em]
      \star &\star & -P\\[0.3em]
     \end{bmatrix}\prec 0
$$
which, up to a congruence transformation, is equivalent to \eqref{eq:M_MI_2} with $S_2=0$. Hence, the result is established. \QEDA
\end{pf}
\end{document}